\documentclass[prb,amssymb, twocolumn,print]{revtex4}

\usepackage{multirow}
\usepackage{pstricks}
\usepackage{graphicx,float}
\usepackage{graphics}
\usepackage{bm}
\usepackage[fleqn]{amsmath}
\usepackage{feyn}

\newcommand{\ri}{\textsl{r}}

\begin{document}
\title [\textsf{The relativistic precession of the orbits}]
{\textsf{The relativistic precession of the orbits}}%
\author{\textsf{Maurizio M.} \surname{\textsf{D'Eliseo}}}%
\email{s.elmo@mail.com}
 \affiliation{Osservatorio S.Elmo - Via A.Caccavello 22,  80129 Napoli Italy}%

\begin{abstract}
The relativistic precession can be quickly inferred
from the nonlinear polar orbit equation without actually solving it.\\
\\
\textbf{Pacs:} 04.20.Cv; 04.25.-g; 04.25.Nx.
\end{abstract}
\maketitle

\section{Introduction}

The precession, and consequently the secular motion of the
pericenter of a bound orbit, stems from the general relativistic
treatment of the motion of a test body in the space-time of a
spherically symmetric distribution of mass which, if also spins,
drags space-time around with it slightly perturbing the orbit
(Lense-Thirring gravitomagnetic force), and then by providing a
further contribution to the precession. The study of the former
phenomenon has been done for various astronomical scenarios. Within
the solar system, have been considered Earth's
satellites,\cite{1,2,3,4,5,6,7,8,9,10,11,12} the Moon\cite{13,14},
Mars,\cite{15} the giant planets,\cite{16,17,18,19,20} the Sun and
its planets.\cite{21,22,23,24,25,26,27} The 1PN post-Newtonian,
Schwarzschild-like orbital effects\cite{28} recently revamped
because of several attempts to detect them in different natural
systems as the galactic environment for stellar orbits around the
central black holes.\cite{29,30,31} Still in a planetary setting,
the exoplanets may constitute a fruitful field of
study.\cite{32,33,34,35,36,37,38,39}

The geodetic equation in the Schwarzschild space-time becomes a
Binet-type differential equation which describes in polar form the
shape of the orbit of a test body under the effect of a force
inversely proportional to the square of the distance from the origin
with added a very small inverse fourth power term. There is no
difficulty in principle, using techniques of perturbation theory, to
deduce the precession by solving the equation for bound orbits to
arbitrary degrees of approximation.\cite{40,41,77,88} Our aim here
is to show a new way to obtain the leading term of the precession in
two moves, consisting of a substitution followed by an elementary
definite integration but, to be able to appreciate its correctness,
it needs to be opportunely introduced and justified.

\section{The angular period operator}

We start from the classical equation of the unperturbed orbit in
polar coordinates\cite{42}
\begin{equation}\label{b1}
u''+u=A,
\end{equation}
where the primes denote double differentiation with respect to the
angle $\phi$. The function $u=u(\phi)$ is the inverse of the radius
vector $r$, while
\begin{equation}\label{l}
A\equiv\frac{\mu u^2}{\ell^2 u^2}=\frac{\mu}{\ell^2}>0,
\end{equation}
where $\mu$ is the gravitational parameter of the central body (this
is strictly correct only when the orbiting body is considered to be
a test-body) and $\ell$ is the area constant. Equation~\eqref{b1}
admits $u=A$ as particular integral, and its complete solution can
be found solving the homogeneous equation $(u-A)''+(u-A)=0$. The
solution in the perifocal system can be written in the form
\begin{equation}\label{e1}
u=A+Ae\cos\phi,
\end{equation}
which is the equation of the elliptical orbit when
$e\,\varepsilon\,(0,1)$. In this setting the angle $\phi$ is the
true anomaly while, in terms of the elliptical elements
$A=u(\pi/2)=1/a(1-e^2)$, where $a,e$ are the semimajor axis and
 the eccentricity. So $1/A$ is the semi-latus rectum and determines the size of the orbit. The fixed position of pericenter is at $\phi=0$, where
$u=1/\ri$ assumes its maximum value. The solution $u$ is periodic of
period $2\pi$, and its average value is
\begin{align}\label{av}
\bar{u}=\frac{1}{2\pi}\int_{0}^{2\pi}u\,d\phi=A,
\end{align}
which is precisely the particular integral of Eq.~\eqref{b1}. It
corresponds to the circular orbit of a body with the given angular
momentum, so that we may associate to the elliptic orbit this
particular circular orbit. They of course share the
$2\pi$-periodicity, and this means that, considering them together,
if we assume that the two radii vectors overlap at $\phi=0$, the
pericenter of the elliptic orbit, they will overlap again at the
successive pericenter $\phi=2\pi$, despite from dynamical point of
view the respective polar angles assume over time different values
in the other positions (apocenter excluded) in a way dictated by the
constancy of the respective areas swept out. But, putting aside the
time which is extraneous to the mathematics of our problem, it is
sufficient to think that the two orbits are connected to one another
as specified only at this special points, and that therefore in
passing from one pericenter to the successive, both radii vectors
rotate through an angle $2\pi$ about the origin. This angle may be
obtained directly from the solution swapping $A$ and $2\pi$ in
Eq.~\eqref{av}, so that we may write
\begin{align}\label{sw}
\hat{P}u\equiv\frac{1}{A}\int_{0}^{2\pi}u\,d\phi&=2\pi,
\end{align}
because the definite integration may be considered as the
application of a linear operator $\hat{P}$ to the solution $u$,
being $\hat{P}$ a mathematical device acting as a detector which
measures the angular period, that is the angle separating two
successive passages of the body through pericenters.

\section{The relativistic precession}

The general relativistic corrections to orbital dynamics imply that
the classical Eq.~\eqref{b1} is substituted by the nonlinear polar
orbit equation\cite{40}
\begin{align}\label{bt}
u''_{r}+u_{r}=A+Bu^2_{r},
\end{align}
 where
$B=3\alpha$, being $\alpha\equiv3\mu/c^2$ the gravitational radius
of the central attractor, generally a tiny fraction of typical
orbital dimensions. Without any attempt to try finding an
approximate bound solution, let's handle it in a simple, but
meaningful way. To do this, wanting to highlight a secular effect on
the orbit, we put $u_{r}^2=\bar{u}^2=A^2 $ in the last term of
Eq.~\eqref{bt}. This means assuming as first approximation to the
solution just that circular orbit we associated to the unperturbed
elliptic orbit, with the shared property with the latter we alluded
to before. We obtain so a linear equation containing a constant
perturbation
\begin{align}\label{ct}
u''_{r}+u_{r}=A_{r},\quad A_{r}\equiv A+A^2B,
\end{align} whose solution in the
perifocal system may be written in the form
\begin{align}\nonumber
u_{r}&=A_{r}+A_{r}\,e\cos\phi=u+A^2B+A^2B\,
e\cos\phi\\\label{up}&=u+\delta u,
\end{align}
namely as the sum of the unperturbed elliptic solution with added a
perturbation term factored by $B$. It is worthy noting that $e$ is
unaffected by the perturbation, so it is still completely arbitrary
in its interval of definition. As we consider Eq.~\eqref{ct}  as a
perturbed version of Eq.~\eqref{b1}, it is legitimate to apply to
the solution~\eqref{up} the operator $\hat{P}$, the detector of the
angular period, and thus we get
\begin{align}\label{co}
\check{P}u_{r}&=\hat{P}u+\hat{P}\delta u=2\pi+2\pi AB,
\end{align}
which also represents the lowest-order contribution to the
precession of the circular orbit per revolution that occurs in the
direction of increasing true anomaly. In substance, the relativistic
term makes a circumference of the circular orbit shorter by $AB$
which in turn makes a deficit of angle of rotation of $2\pi AB$. So,
to complete one revolution along the orbit of radius $1/A$ for which
is calibrated the operator $\hat{P}$, the orbiting body needs to
rotate this same extra angle and this in turn is reflected in the
pericenter shift of the associate elliptic orbit because of the
compulsory coincidence of the angular positions of the two radii
vectors at this special point. In conclusion, expressing $A$ in
terms of the elliptical elements of the orbit, from Eq.~\eqref{co}
it follows the Einstein's precession formula
\begin{align}\label{po}
\hat{P}u_{r}=2\pi+\frac{6\pi\alpha}{a(1-e^2)}.
\end{align}
The essential feature of the approach followed (a particular
averaging method) is that, without the need of using any specific
perturbation equation, it employs the circular motion as a test
motion for which, thorough the operator $\hat{P}$, is easy to obtain
the leading precession term, which is \emph{pari passu} transferred
to the elliptical orbit. When applied to other types of
inverse-power perturbing central forces (not necessarily of academic
interest) this method gives the dimensionally correct algebraic form
of the precessional term but off target by a small numerical factor,
so underestimating the proper orbital precession determined with
other methods.\cite{43,44,45,46,47,48,49,50,51,52,53} We deduce that
in these instances the procedure cannot fully capture the real
variation of $A$ on the sole basis of the unperturbed averaged value
of $u$ employed in the approximation, and this is perhaps due to the
fact that the constant $A$ appears linearly after the action of the
operator $\hat{P}$ only with the occurrence of a quadratic
nonlinearity as in Eq.~\eqref{bt}, and we were lucky enough that in
the most important case of physical interest it works well. In the
other cases, the method should be refined.

\section{Conclusions}

It would be interesting to see whether it can similarly treat
another case of possible relativistic precession, the Lense Thirring
drag,\cite{54} to which we dedicate here only a fleeting mention.
This effect closely parallels the way in which a rotating
electrically charged body generates magnetism. The analogy has
helped to organize the understanding of the phenomenon and to
determine predictions experimentally testable. The Gravity Probe B
satellite\cite{55,56} achieved an accuracy within 19 per cent of the
expected orbit change; other satellites got a rather similar level
of accuracy. Researchers hope to achieve 1 per cent with
LARES.\cite{57,58,59} In this phenomenological approach the forces
in play are non-central, so that our method is not directly
applicable, but it would not be difficult to conceive a model in an
appropriate plane of symmetry which allows to simply deduce, with
suitable geometrical and physical insight to justify the mathematics
employed, the leading precessional term with a minimum effort. We
suggest also to use the method in some non-Newtonian model of
gravity, as the Yukawa-like fifth force.\cite{60} Here we have no
problems of non-centrality, so it is easy write down the relative
polar equation and hence the precession term, but after that
assumptions must be made on the figures to be assigned to the free
parameters in such a way as to obtain testable results for realistic
models of astrophysical systems.

\end{document}